\documentclass[proof]{pasj00}
\usepackage{ulem}

\begin{document}
\SetRunningHead{Author(s) in page-head}{Running Head}
\Received{2009 August 17}
\Accepted{2009 November 5}

\title{Study of the Large-scale Temperature Structure of the Perseus
Cluster with Suzaku}

\author{Sho NISHINO$^{1}$, Yasushi FUKAZAWA$^{1}$,
Katsuhiro HAYASHI$^{1}$,\\Kazuhiro NAKAZAWA$^{2}$, Takaaki TANAKA$^{3}$\\
{\small $^{1}$ Department of Physical Science, Hiroshima University, 1-3-1 Kagamiyama,}\\
{\small Higashi-Hiroshima, Hiroshima 739-8526, Japan}\\
{\small $^{2}$ Department of Physics, University of Tokyo, 7-3-1 Hongo,
Bunkyo, Tokyo 113-0033, Japan}\\
{\small $^{3}$ Kavli Institute for Particle Astrophysics and Cosmology, Stanford  
University, 2575 Sand Hill Road M/S 29, Menlo Park, CA 94025, USA}}


%


\maketitle

\begin{abstract}
We report on a study of the large-scale temperature structure of the
Perseus cluster with Suzaku, using the observational data of 
four pointings of 30' offset regions, together with the data from 
the central region. 
Thanks to the Hard X-ray Detector (HXD-PIN: 10 - 60 keV),
Suzaku can determine the temperature of hot galaxy clusters.
We performed the spectral analysis, by considering the temperature
structure and the collimator response of the PIN correctly.
As a result, we found that
the upper limit of the temperature in the outer region is $\sim$ 14 keV, and an
extremely hot gas, which was reported for RXJ 1347.5-1145 and A 3667, 
was not found in the Perseus cluster.
This indicates that the Perseus cluster has not recently experienced a major
merger.
\end{abstract}

\section{Introduction}

Clusters of galaxies are widely believed to grow up with repetition of
small-scale cluster merging. Gravitational energy released in the 
merger process
heats up intracluster plasma, and accelerates particles up to higher energy. 
Actually, in many clusters, diffuse synchrotron radiation from high
energy electrons is detected by radio observations (e.g., A2163: Feretti et
al. 2001). In addition, a sign of 
the hard X-ray emission, that is thought to be Inverse Compton
Scattering (ICS) of the Cosmic Microwave Background by accelerated
electrons, has been reported for some clusters 
(e.g., Coma cluster: Fusco-Femiano et al. 1999).
Moreover, large temperature fluctuations are often found in clusters
which are candidates of non-thermal hard X-ray emitters.
Recent X-ray studies have reported that very hot plasma, whose
temperature exceeds $\sim 20$ keV, exists in some merging clusters, 
such as RX J1347.5-1145 (Ota et al,
2008) and A 3667 (Nakazawa et al, 2009). 
As above, plasma heating and particle acceleration in clusters are inseparably
connected phenomena. Therefore, search for these phenomena is
important to understand the history of cluster evolution, 
especially for the physical mechanisms of plasma heating and energy transportation.

In this paper, we report on a study of the large-scale temperature
structure of the Perseus cluster (Abell 426). The Perseus cluster is a
nearby (z = 0.0183), massive, largely extended cluster, and is the most
luminous cluster in the X-ray band. 
An X-ray bright active galaxy NGC 1275, with a radio
mini-halo, is located at the cluster center, and non-thermal
power-law emission from NGC 1275 was confirmed by past 
observations (e.g., Sanders et al, 2005).     
ASCA found a large fluctuation of temperature
in this cluster, and indicated that a very hot region with the temperature
exceeding $\sim 10$ keV exists
in the outer region. Information on cluster merging should
remain in the low-density outer rather than the dense
central region. Therefore, it is very valuable to investigate the temperature
structure and non-thermal emission in the outer region of
the cluster carefully.

Determination of the temperature $kT$ of hot regions with $kT\geq 10$ keV
is difficult for detectors, whose energy band is limited below 10 keV 
(e.g. ASCA and XMM-Newton).
Therefore, we should observe clusters with a detector, that is sensitive
above 10 keV, so as to determine the temperature by covering 
the spectral roll-off of thermal emission.
Therefore, we observed 30' offset regions from the Perseus cluster center,
with the HXD-PIN/XIS onboard Suzaku.
HXD-PIN is a non-imaging detector of 64 PIN diodes, covering the hard X-ray
band of 10-60 keV, and able to perform observations with a low background
level and a small field of view (FOV) of 34'$\times$34' (FWHM).
A narrow field of view has advantage of reducing the contribution of the
bright central region to the observed spectra.
The XIS is a focal plane CCD detector with an X-ray Telescope (XRT), covering
the soft X-ray band of 0.2-10 keV with an FOV of 18'$\times$18'.
Combination of the XIS and HXD-PIN gives a broad band X-ray spectrum and
thus we can determine the temperature structure of hot clusters.
Throughout this paper, we adopt a Hubble constant of H$_{0}$ = 50
h$_{50}$ km s$^{-1}$ Mpc$^{-1}$. All statistical errors are given at 90\%
confidence level.

\section{Observation and Data Reduction}

\begin{figure}[!h]
\begin{center}
\includegraphics*[scale=0.55]{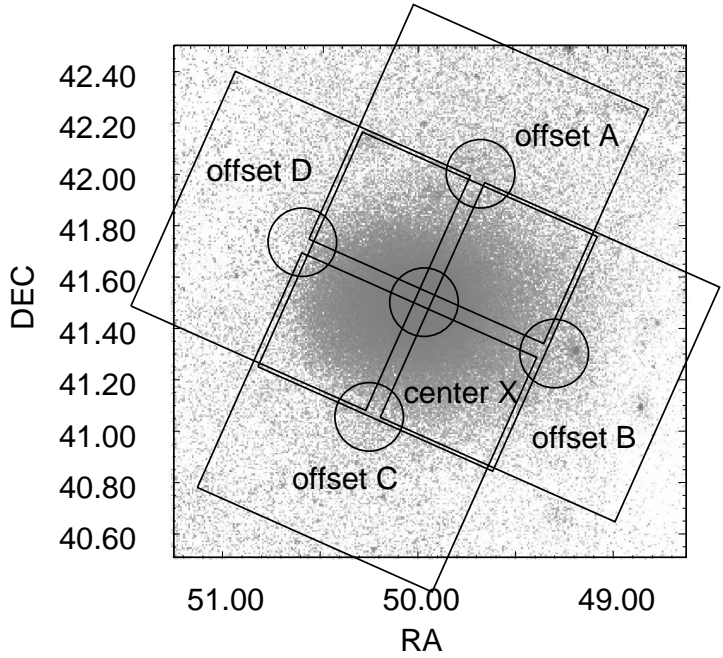}
\end{center}
\vspace{-0.5cm}
\caption{2 deg $\times$ 2 deg image of the Perseus cluster obtained 
with the ROSAT-PSPC (0.5-2  keV). The FOV of the center observation X and the
 offset observations A,B,C,D are shown on the image. 
Box shapes are the FOV of HXD-PIN ($1^{\circ} \times 1^{\circ}$), and circles are
 the FOV of XIS (8'-radius).}
\label{fov}
\end{figure}

We observed the outer regions of the Perseus cluster on September 2-4, 2006,
with Suzaku. Four pointing observations of 30' offset regions from the 
cluster center were carried out. 
These observations make use of a narrow FOV of the PIN, 
by reducing the contribution of intense emission from the
cluster center.
Additionally, we also analyzed the Suzaku public data of observations of
the Perseus cluster center, on 
February 1--2, 2006, August 29 -- September 2, 2006, and February
5--6, 2007. Fig \ref{fov} and Table \ref{obs} summarize these observations.

\begin{table}[!h]
\caption{Summary of Suzaku observations of the Perseus cluster for our analysis.}
\label{obs}
\begin{center}
\begin{tabular}{lllll}\hline \hline
Position & Sequence No. & Date & RA, DEC & Exposure (HXD/XIS) \\ \hline
Center X & 800010010    & 2006/02/01-02  & 49$^\circ$.9504 / 41$^\circ$.5117  &52ks/41ks \\ 
         & 101012010$\star$    & 2006/08/29-/09/02  & 49$^\circ$.9504 / 41$^\circ$.5117  &129ks/- \\
         & 101012020$\star$    & 2007/02/05-06  & 49$^\circ$.9504 / 41$^\circ$.5117  &41ks/- \\ 
Offset A & 801049010    & 2006/09/02-02  & 49$^\circ$.6833 / 42$^\circ$.0081  &23ks/23ks \\  
Offset B & 801049020    & 2006/09/02-03  & 49$^\circ$.3167 / 41$^\circ$.3131  &24ks/24ks \\  
Offset C & 801049030    & 2006/09/03-04  & 50$^\circ$.2625 / 41$^\circ$.0411  &29ks/29ks \\
Offset D & 801049040    & 2006/09/04-04  & 50$^\circ$.6125 / 41$^\circ$.7464  &6ks/6ks\\ \hline
\multicolumn{5}{l}{$\star$Data of 101012010 and 101012020 were used only for the spectral analysis of the HXD-PIN.}
\end{tabular}
\end{center}
\end{table}

We used the version 2.0 pipeline processing data, and data screening was
performed with HEAsoft ver 6.2. 
HXD-PIN data were screened with a cut-off
rigidity (COR) of $>$ 6 GV, elevation angle of $>$ 5$^{o}$ from
the Earth rim, and good time intervals (GTI) during which the satellite is
outside the South Atlantic Anomaly (SAA).
We used ae\_hxd\_pinhxnome\{1,2,3\}\_20080129.rsp for the HXD
response matrix, and a "tuned" background (bgd-d) for the non X-ray background
(NXB) (Fukazawa et al. 2009)
of the PIN detector. Both were supplied by the HXD team (Fukazawa et
al. 2008). 
The cosmic X-ray background (CXB) contribution for the HXD data was estimated 
by using the CXB parameters 
of Kirsch et al.(2005) and considering the collimator response, 
and then subtracted from the observed spectra as well as the NXB.

Data screening of the XIS data was almost the same as for the HXD,
except for applying the criteria of COR $>$ 8 GV and elevation
angle of $>$ 20$^{o}$ from the Earth rim. 
Response matrix files (rmf) and auxiliary response files (arf) were
generated by the FTOOL xisrmfgen and xissimarfgen (Ishisaki et al. 2007), 
respectively, and the NXB was estimated by the FTOOL xisntebgdgen (Tawa et al. 2008).

\section{Analysis and Results}

Observed PIN spectra contain photons coming from a large sky area of the
cluster, due to its FOV of $34'\times34'$ (FWHM). Therefore, we have to take
into account this effect to determine the temperature structure of the cluster.
On the other hand, the XIS can obtain the spectrum of small regions, while it
cannot determine the temperature of hot gas above 10 keV accurately, 
because of the limitations of the energy band below 10 keV.
In the following analysis, we applied the APEC model to represent 
thermal emission from the intracluster plasma. 
Furthermore, we multiplied the plasma model by the WABS model, to
take account of the photoelectric absorption in the Galactic intersteller
medium. Projection effects in the line of sight are not considered here.

\subsection{Nuclear emission from NGC 1275}

Nuclear X-ray emission from NGC 1275, which is located at
the cluster center, has been reported by past studies, for
example, XMM-Newton (Churazov et al. 2003), Chandra (Sanders et al. 2005)
and Swift/BAT (Ajello et al. 2008).
Eckert et al. (2009) showed that high-energy flux ($>$30 keV) is variable
over a time scale of several months with INTEGRAL data, and proved that
the origin of the high-energy emission cannot be diffuse.
Therefore, we have to estimate the X-ray flux of the nucleus as contamination, 
to obtain the accurate temperature structure.

\begin{figure}[!h]
\begin{center}
\begin{minipage}{5cm}
\vspace{0.2cm}
\includegraphics[width=4.2cm,clip]{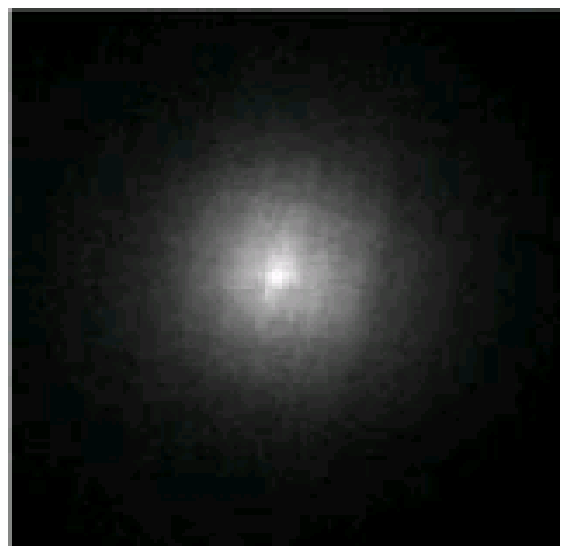}
\end{minipage}
\hspace{0cm}
\begin{minipage}{5cm}
\vspace{0.2cm}
\includegraphics[width=4.2cm,clip]{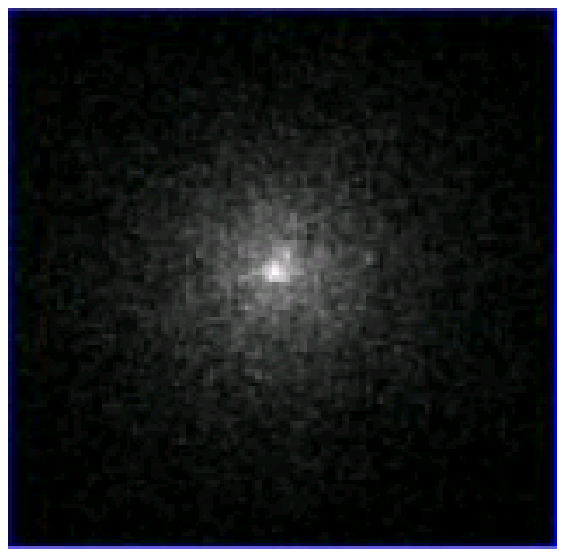}
\end{minipage}\\
\begin{minipage}{5cm}
\vspace{0.2cm}
\hspace{-0.5cm}
\includegraphics[width=5.cm,clip]{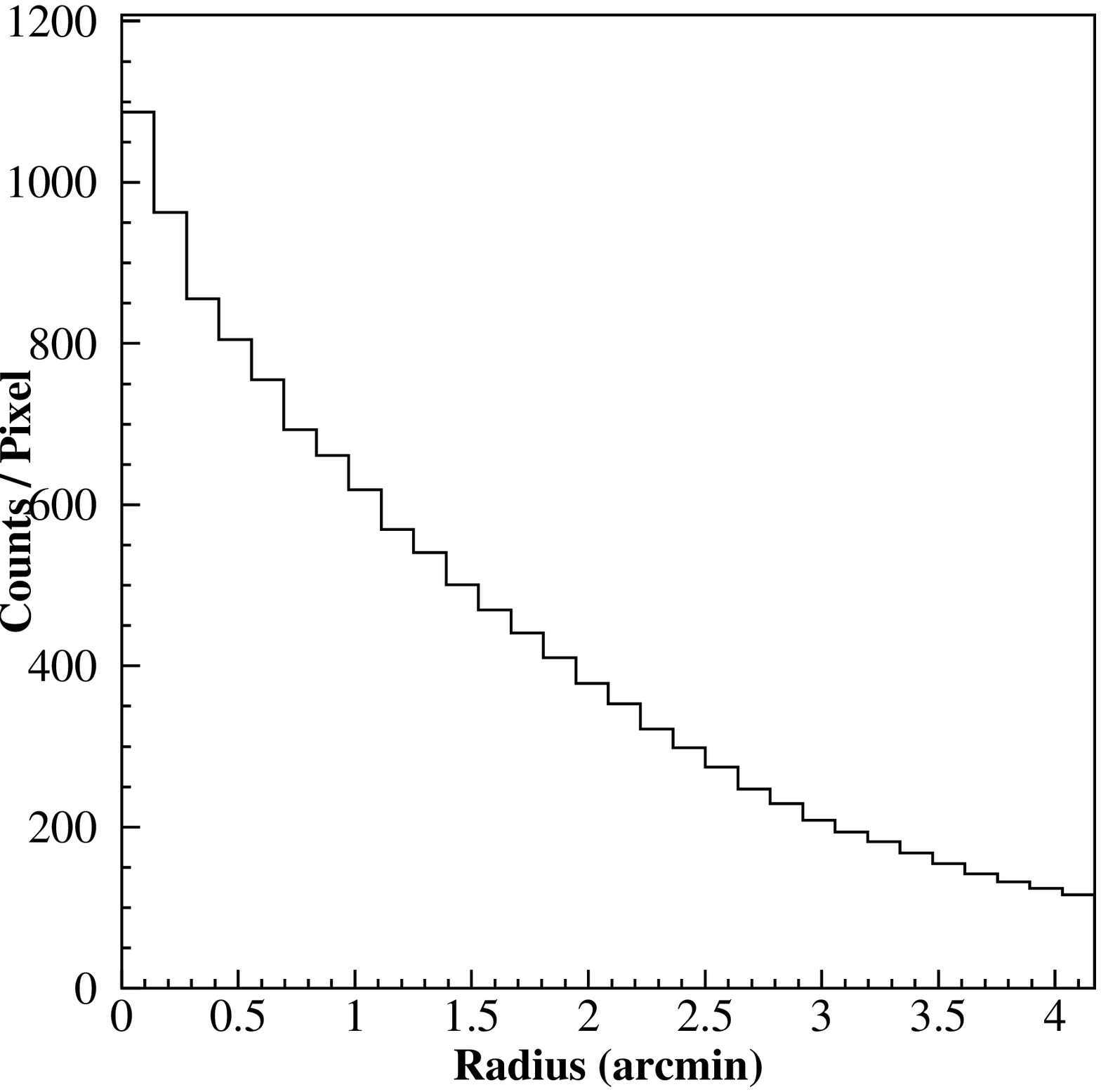}
\end{minipage}
\hspace{0cm}
\begin{minipage}{5cm}
\vspace{0.2cm}
\hspace{-0.5cm}
\includegraphics[width=5cm,clip]{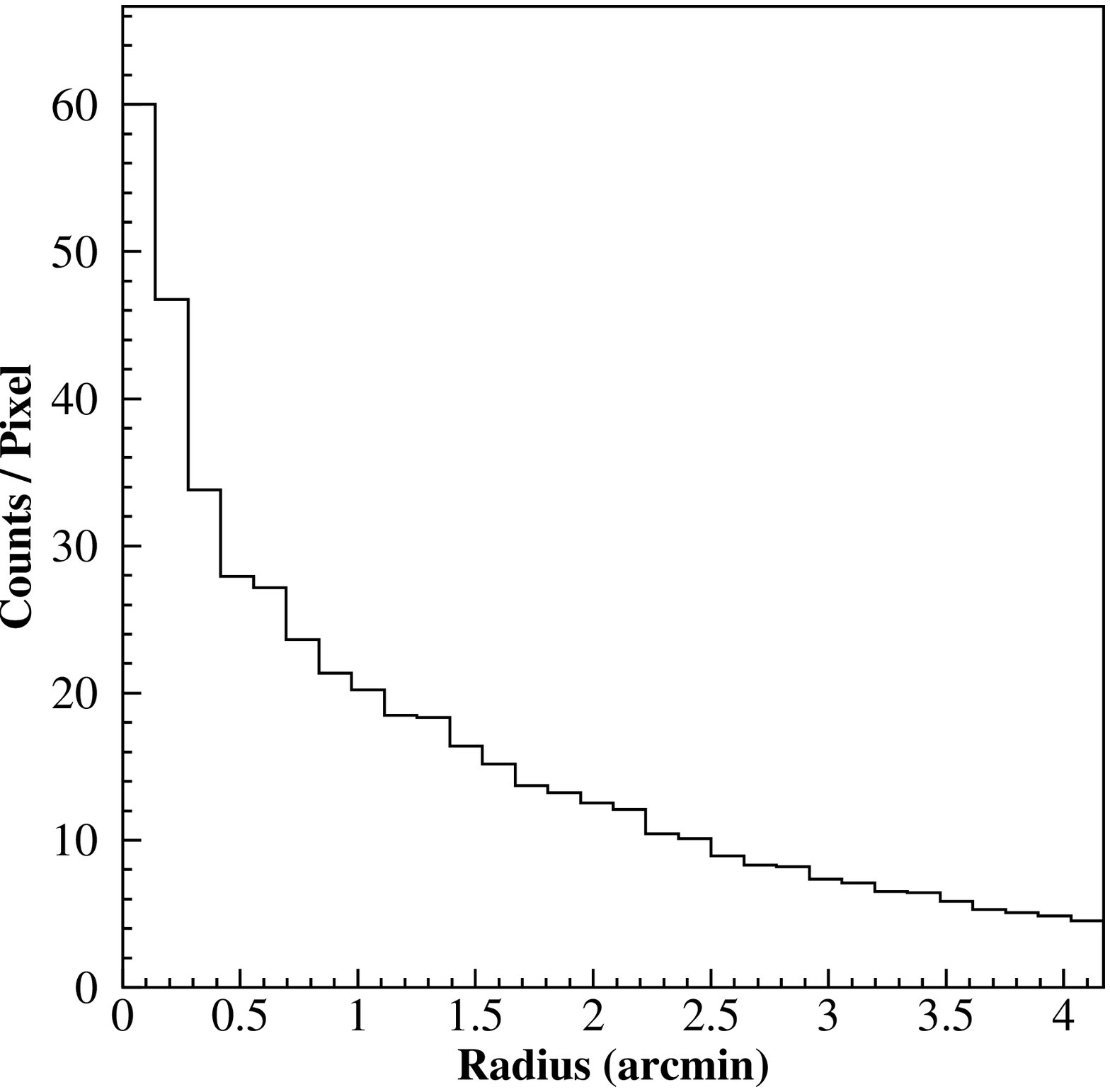}
\end{minipage}
\end{center}
\caption{Top panels are XIS images (18'$\times$18') extracted in the 4-5
 keV (left) and 9-10 keV (right) band. Bottom panels are radial surface 
brightness distributions extracted in the 4-5 keV (left) and 9-10 keV
 (right) band.}
\label{pror}
\end{figure}

\begin{figure}[!h]
\begin{center}
\rotatebox{-90}{\includegraphics[scale=0.4]{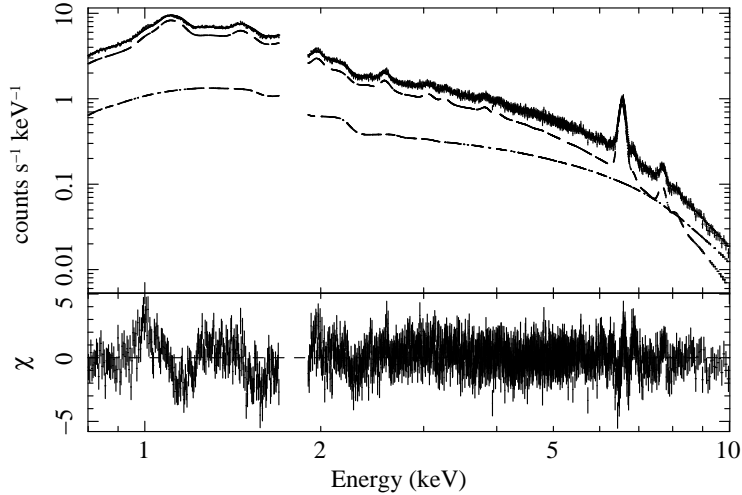}}
\end{center}
\caption{XIS spectrum within 2' radius with the best-fit model WABS $\times$
 (APEC
 + POWERLAW). Dashed lines in the upper panel show the APEC
 and POWERLAW contributions,  respectively, and the residuals are shown in the bottom panel.}
\label{hspec}
\end{figure}

Fig \ref{pror} shows XIS images and radial surface brightness distributions
in the 4-5 keV and 9-10 keV band. The central excess is
clearly confirmed in the XIS image within 1' radius, especially in the
hard X-ray band. 
Taking into account the point spread function of the XRT/XIS, this central
excess is consistent with a point source, that is, nuclear emission 
of NGC 1275.
We performed spectral fitting for the region within 2' radius, with a
single temperature APEC plus POWERLAW model. 
We generated the arf with xissimarfgen by using the XIS image within 2
arcmin from the cluster center as a seed image.
Here, the photon index of
the POWERLAW model is a free parameter. 
We show the spectrum and best-fit parameters in Fig \ref{hspec}
and Table \ref{hpara}, respectively. 
The fit improves from $\chi^{2}$/d.o.f =
2.38 to 1.31, when adding the POWERLAW model to the single 
temperature APEC model.
The estimated power-law luminosity is $\sim 5.6 \times 10^{43}$ erg s$^{-1}$
(0.8-10 keV), in agreement with the 
value reported by Chandra (Sanders et al. 2005) of $\sim 6 \times
10^{43}$ erg s$^{-1}$ (within 3' radius).

\begin{table}[!h]
\footnotesize
\caption{Fitting results of the XIS spectrum within 2' radius with
 a model WABS $\times$ (APEC+POWERLAW).}
\label{hpara}
\begin{center}
\hspace*{-1.0 cm}
\begin{tabular}{lllllll} \hline \hline
$N_{\rm H}$ & kT &abundance & photon index & reduced $\chi^{2}$ &  PL flux(0.8-10
 keV) & total flux(0.8-10 keV)     \\ \hline
($\times 10 ^{21}$ cm$^{-2}$)  &(keV) & (solar) & (-) &
 ($\chi^{2}$/d.o.f)& (erg cm$^{-2}$ s$^{-1}$) & (erg cm$^{-2}$ s$^{-1}$)\\ \hline
1.32 $\pm$0.01  &3.25$\pm$0.02  & 0.70$\pm$0.01 & 1.61$\pm$0.02 & 2048/1569 & 8.27$\pm 0.06 \times 10^{-11}$ & $2.92 \pm 0.01 \times 10 ^{-10}$ \\ \hline
\end{tabular}
\end{center}
\end{table}

\subsection{Overall Properties}

\begin{figure}
\begin{center}
\includegraphics*[scale=0.6]{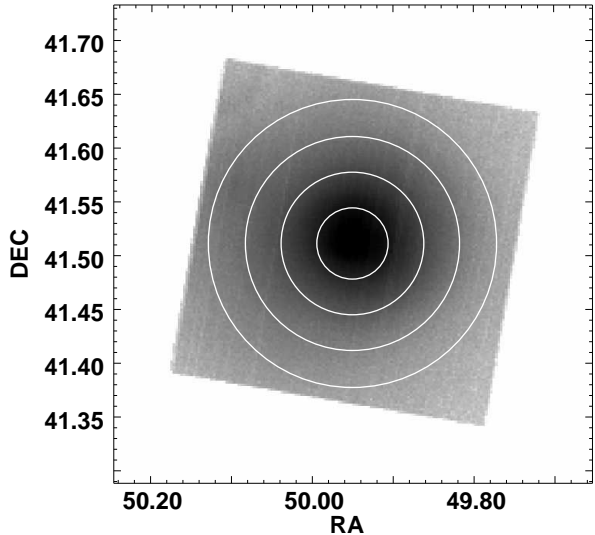}
\end{center}
\caption{XIS image of the center observation. The radii of the white circles
 are 2', 4', 6', 8' from the cluster center.}
\label{centerX} 
\end{figure}

In this section, as a preliminary step prior to the detailed analysis, we
investigate the overall properties of the cluster temperature structure. 
At first, we analyzed the XIS data of the cluster center to obtain the
temperature structure in the central region.   
We divided the XIS events into 4 annular regions of $0'-2'$, $2'-4'$, $4'-6'$,
$6'-8'$ in radius from the cluster center (Fig \ref{centerX}).
The result of $0'-2'$ was given in the previous subsection.
For the regions of $2'-4'$, $4'-6'$, $6'-8'$, we performed spectral fitting
with a single temperature APEC model. 
We generated the arf of each region with xissimarfgen by using the XIS image
extracted in each spectral region as a seed image.
Here, each region was assumed to be isothermal.
Best-fit parameters for each region are summarized in Table \ref{cx}.
The total flux of the 4 annular regions is 
approximately $1 \times 10^{-9}$ erg cm$^{-2}$ s$^{-1}$ (0.8--10 keV), and
consistent with the past results of ASCA (Fukazawa et al. 2004).
The obtained temperature decrement and
abundance increment toward the center are roughly
consistent with the past results of XMM-Newton (Churazov et al. 2003) and 
ASCA (Ezawa et al. 2001).

\begin{table}[!h]
\caption{Result of fits of XIS spectra in 2'-4', 4'-6', 6'-8' annuli
 with the WABS $\times$ APEC model.}
\label{cx}
\begin{center}
\begin{tabular}{lllllll} \hline \hline
region    & $N_{\rm H}$ & kT  &  abundance & reduced $\chi^{2}$ &flux(0.8-10 keV)\\ \hline
(radius)  & ($\times 10 ^{21}$ cm$^{-2}$)   & (keV) & (solar) &
 ($\chi^{2}$/d.o.f) & (erg cm$^{-2}$ s$^{-1}$) \\ \hline
2' - 4':  & 1.07$\pm$0.01  & 4.60$\pm$0.01  & 0.55$\pm$0.01 & 3559/1696
 & $3.50  \pm 0.01 \times 10^{-10}$ \\          
4' - 6':  & 1.05$\pm$0.01    & 5.88$\pm$0.05  & 0.47$\pm$0.01 & 2018/1480 &
 $2.33  \pm 0.01 \times 10^{-10}$ \\
6' - 8':  & 1.05$\pm$0.01    & 6.45$\pm$0.08  & 0.42$\pm$0.01 & 1439/1158 &
 $1.70  \pm 0.01 \times 10^{-10}$ \\ \hline
\end{tabular}
\end{center}
\end{table}

\begin{figure}[!h]
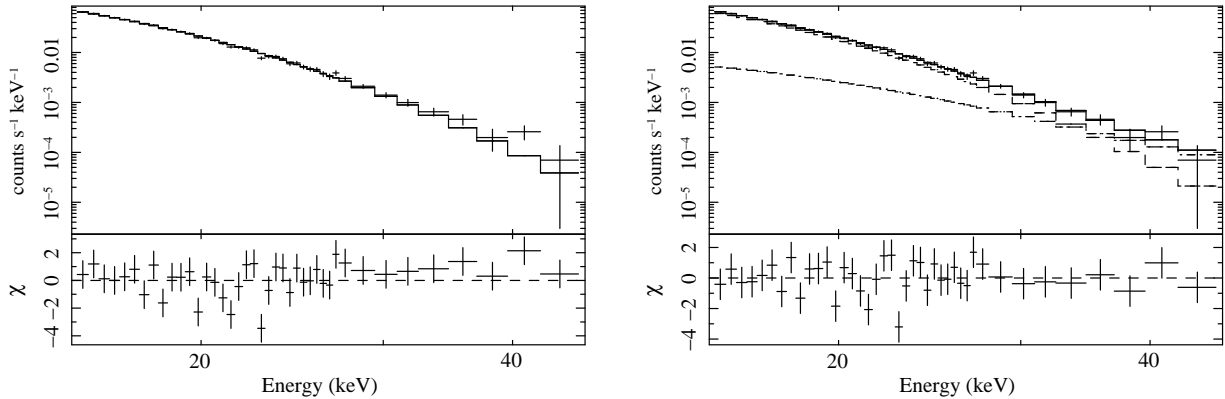

\begin{center}
\begin{minipage}{8cm}
\vspace{0.2cm}
\rotatebox{-90}{\includegraphics[scale=0.32]{./paper_fig5a.eps}}\\
\end{minipage}
\hspace{0.2cm}
\begin{minipage}{8cm}
\vspace{0.2cm}
\rotatebox{-90}{\includegraphics[scale=0.32]{./paper_fig5b.eps}}\\
\end{minipage}
\end{center}
\caption{Background-subtracted HXD-PIN spectrum of the center
 observation with the best-fit model of the single-temperature APEC (left), and
 the single-temperature APEC + POWERLAW (right). Dashed lines in the
 right upper panel are the same as in figure 3.}
\label{centerpinfig}
\end{figure}

Next, we derived the temperature by using the PIN data of the central 
observation. 
The PIN data of the three center observations were added so as to obtain 
enough data statistics.
The arf was generated by the FTOOL hxdarfgen. The current version of
hxdarfgen calculates the arf for a point source.
Therefore, for this analysis, we divided a 2 deg
$\times$ 2 deg cluster image into 100$\times$100 regions, and calculate 
the arf for each region. These arf files are weight-averaged by
intensity, based on the cluster image of the ROSAT-PSPC
\footnote{SkyView: http://skyview.gsfc.nasa.gov/}.
We tried to fit the PIN spectrum with a single temperature APEC
model. Furthermore, 
in order to consider the emission from NGC 1275, 
we also tried to fit with a single temperature APEC plus 
POWERLAW model.
Here, the photon index of the power law was fixed to
1.61, which is obtained by analysis of the XIS data in \S 3.1. The metal
abundance of 0.4 solar is an average value in the center region,
obtained from the XIS spectral analysis.
To avoid the thermal noise of the PIN, we did not use
the data below 15 keV for spectral fitting. 
The fitting results are shown in Table \ref{centerpinpara} and 
Fig \ref{centerpinfig}. 
Without the POWERLAW model, the temperature is $\sim$ 7.2 keV. 
The estimated flux scaled to the total flux of the whole cluster region 
becomes 1.7$\times$10$^{-10}$ erg cm$^{-2}$ s$^{-1}$ (15-50 keV) or
1.8 $\times$10$^{-9}$ erg cm$^{-2}$ s$^{-1}$ (0.8-10 keV), which is 
roughly consistent with the XIS results.
With the POWERLAW,
a temperature of $\sim$ 6.3 keV, and power-law luminosity of $ \sim 2.5
\times 10^{43}$ erg s$^{-1}$ are obtained.

Next, we try to estimate the contribution of the central cool region
to the PIN spectrum. We fitted the average XIS spectra within 8 arcmin
with a single temperature APEC model, and obtained a temperature of
5.5 keV. Then, we fitted the PIN spectrum with a two-temperature APEC
model plus powerlaw model. For one APEC model, the temperature and
normalization are fixed to the values obtained from the above XIS
analysis. Then, the temperature of the hot APEC component becomes 9.5$^{+2.2}_{-1.7}$ keV.
Therefore, the XIS and PIN spectra do not need an extremely hot
component.

\begin{table}[!h]
\footnotesize
\caption{Fitting results of the HXD-PIN spectrum 
of the cluster center observation
 with the model of APEC or APEC+POWERLAW.}
\label{centerpinpara}
\begin{center}
\begin{tabular}{llllll} \hline \hline
kT &abundance & photon index & reduced $\chi^{2}$ &  PL flux(15-50 keV)
 & total flux(15-50 keV)     \\ \hline
(keV) & (solar) & (-) & ($\chi^{2}$/d.o.f)& (erg cm$^{-2}$ s$^{-1}$) &
 (erg cm$^{-2}$ s$^{-1}$) \\ \hline
$7.2 \pm 0.2 $ & 0.4 fixed & -  & 54/39 &
 - & $1.71 \pm 0.03 \times 10^{-10}$  \\
$6.4 \pm 0.5 $ & 0.4 fixed & 1.61 fixed & 42/38 &
 $3.2 \pm 1.3 \times 10^{-11}$ & $1.81 \pm 0.05  \times 10 ^{-10}$ \\ \hline
\end{tabular}
\end{center}
\end{table}

\begin{figure}[!h]
\begin{center}
\begin{minipage}{7cm}
\includegraphics[scale=0.3]{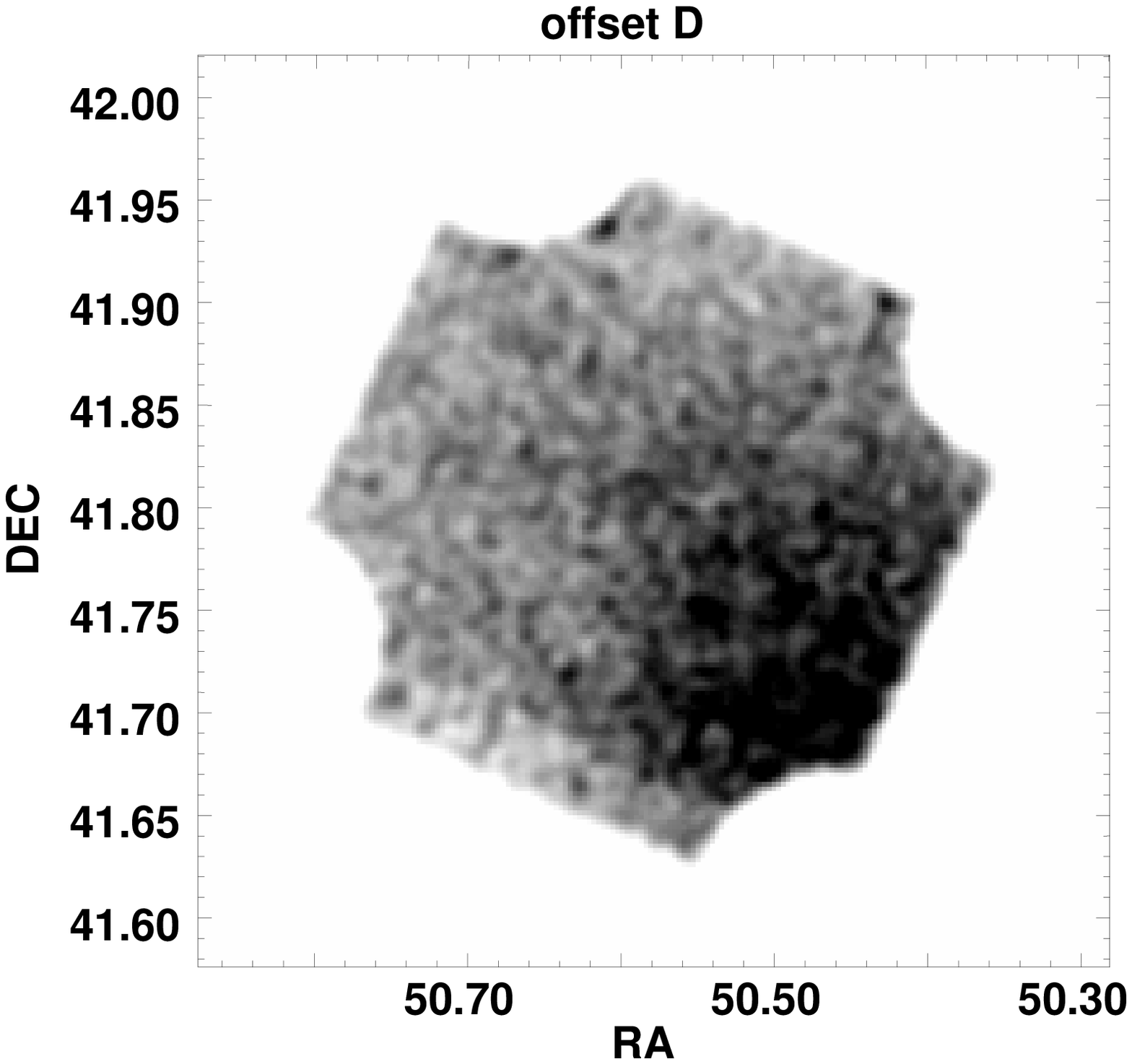}
\end{minipage}
\hspace{-1cm}
\begin{minipage}{7cm}
\includegraphics[scale=0.3]{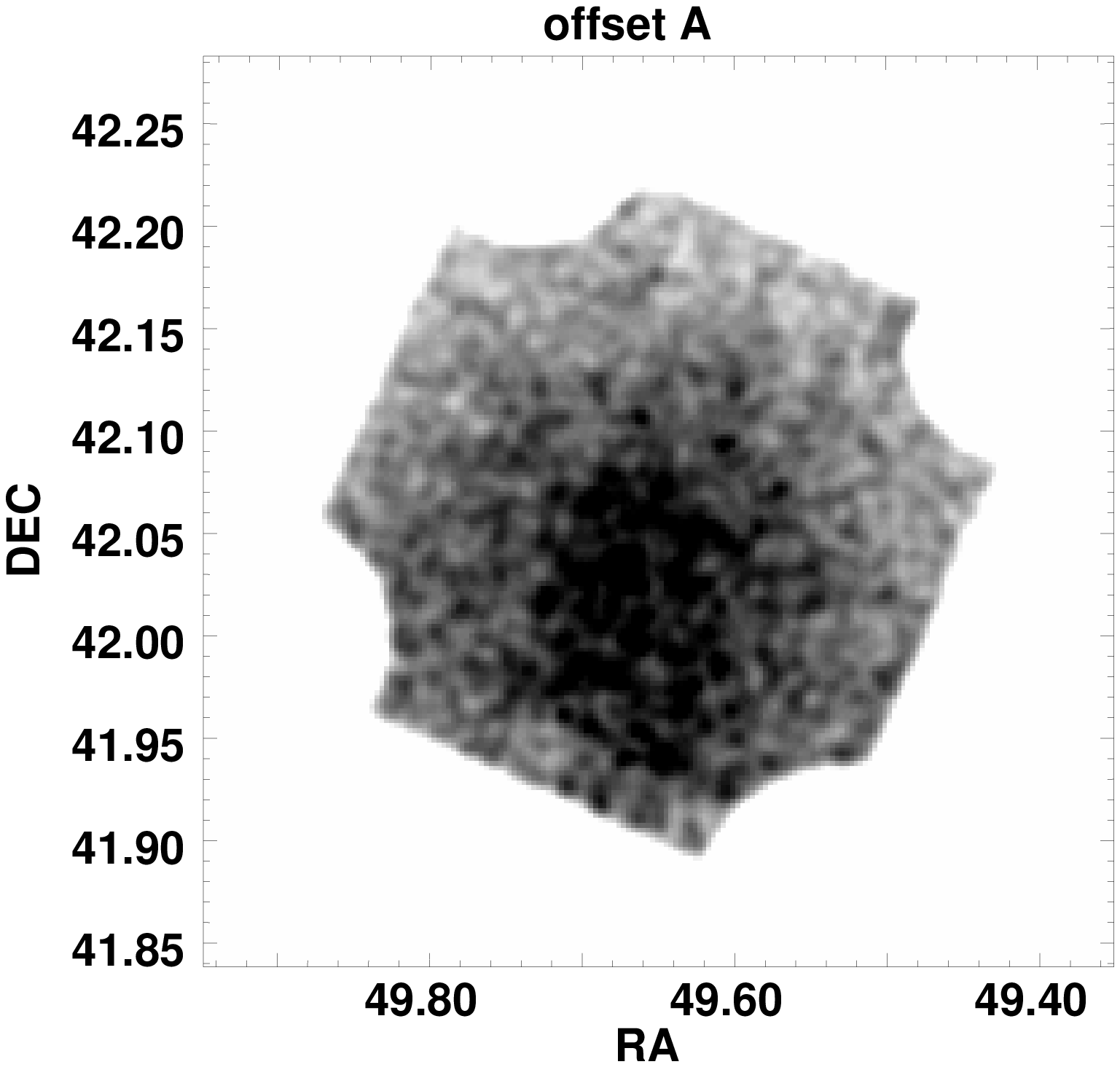}
\end{minipage}\\
\begin{minipage}{7cm}
\vspace{-0.2cm}
\includegraphics[scale=0.3]{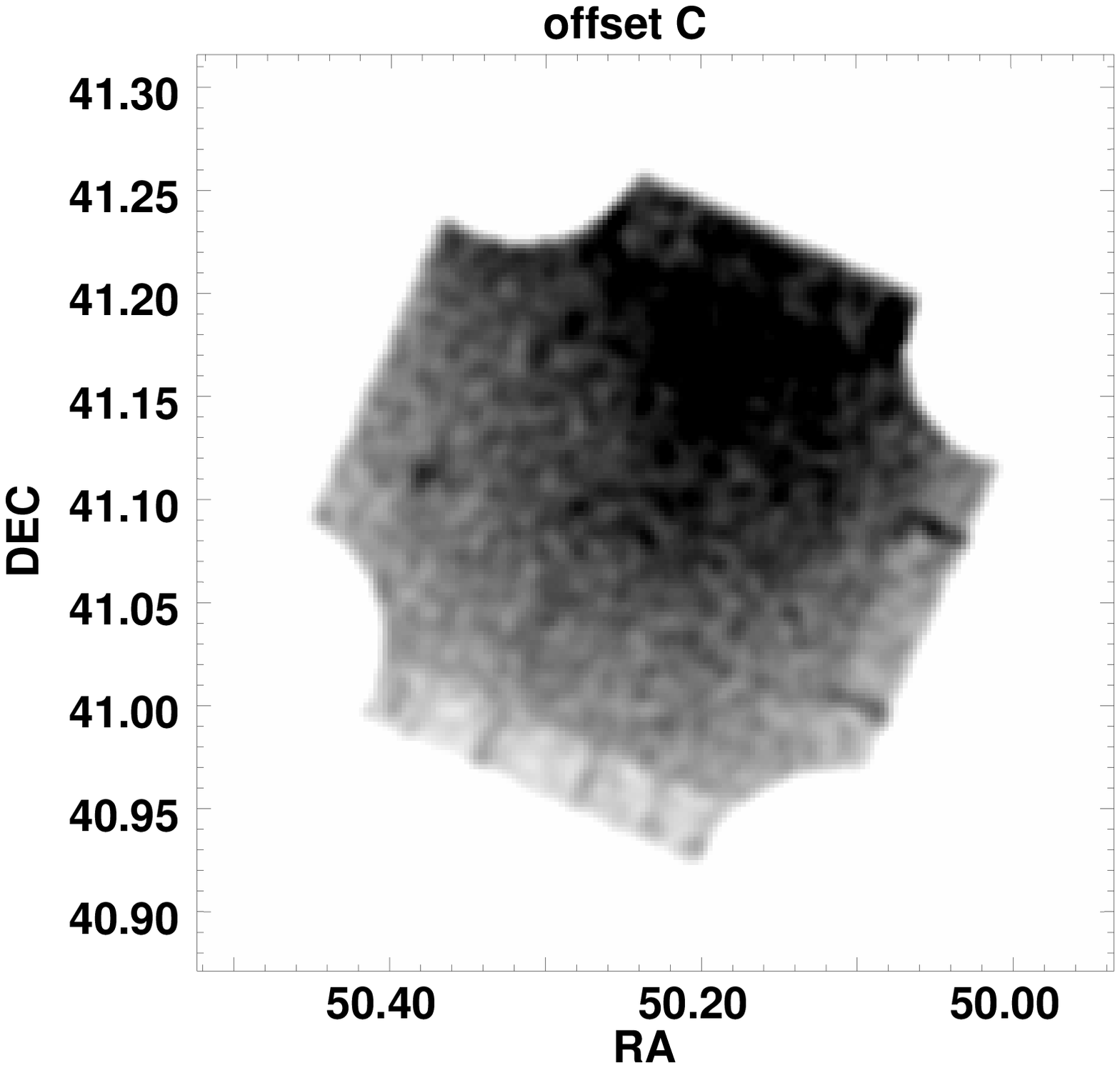}
\end{minipage}
\hspace{-1cm}
\begin{minipage}{7cm}
\vspace{-0.2cm}
\includegraphics[scale=0.3]{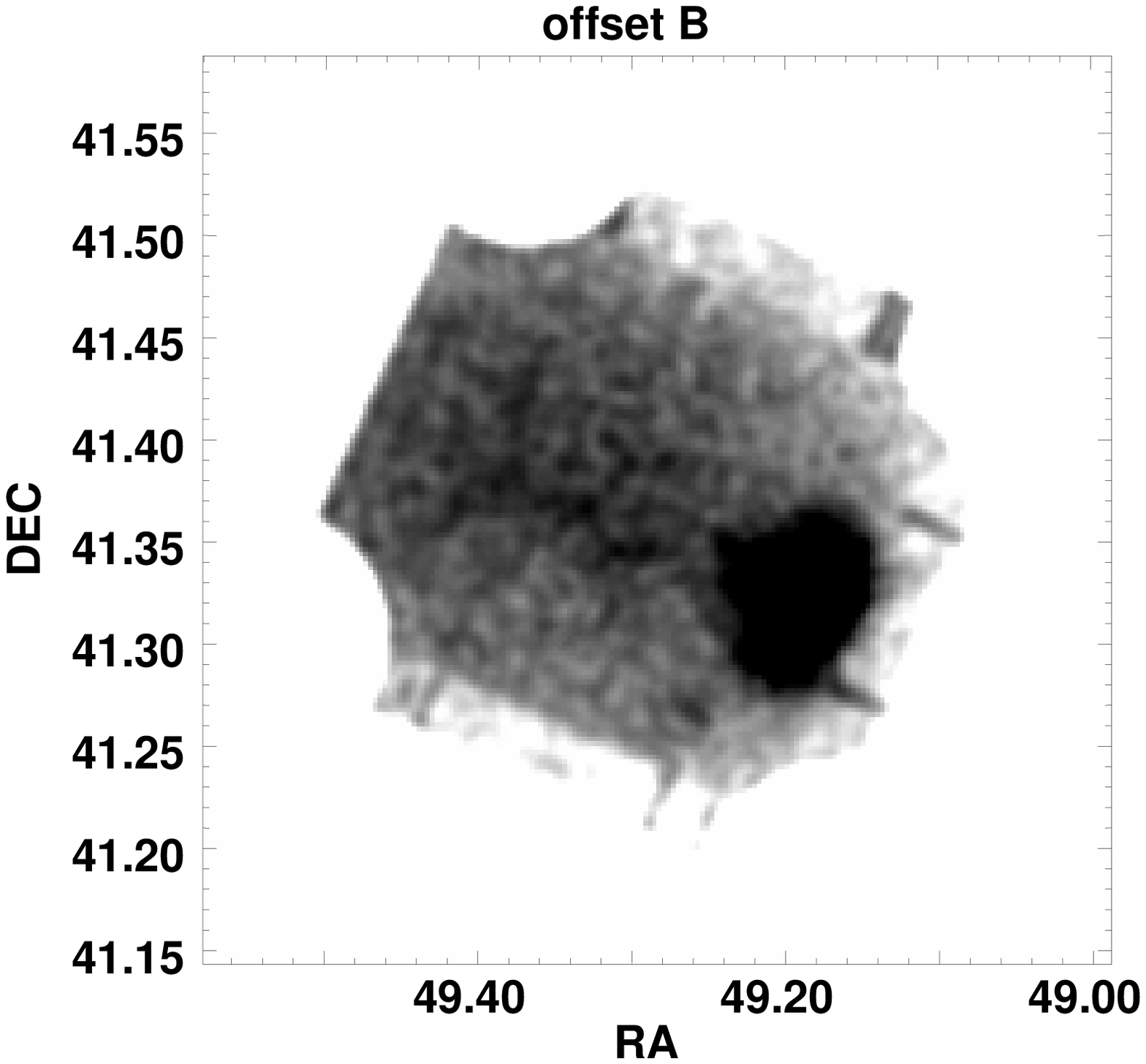}
\end{minipage}
\end{center}
\caption{XIS images of offset A, B, C, D observations. Calibration
 source regions at the corner of each image are subtracted. A point
 source, IC 310, is visible in the FOV of the offset B observation. }
\label{offxis}
\end{figure}

In the same way, we derived the temperature of each offset region, by
using the offset XIS data (Fig \ref{offxis}), with a single temperature model.
The absorption column density is fixed to 1.0 $\times
10^{21}$ cm$^{-2}$, which is the value obtained by the
XIS analysis as shown in Table \ref{cx}. Results are shown in Fig
\ref{offxisfit}. 
The temperature is about 5 - 7 keV in any region,  
and a hot component with $kT>$10 keV is not required for XIS spectra.
If we adopt a higher absorption column density as 1.2
or 1.4$ \times 10^{21}$ cm$^{-2}$, a slightly lower, by less
than $\sim$ 0.5 keV, temperature is required.
The iron abundance is about 0.2-0.3 solar. These results are roughly
consistent with those reported by ASCA (Ezawa et al. 2001).

We used the ROSAT-PSPC image of the Perseus cluster as a seed
image of xissimarfgen, that is,   
fluxes shown in Table \ref{offxisfit} are scaled to the total cluster
flux, taking account of the emission
distribution of the cluster and the vignetting effect of the XRT.
Then, the obtained flux of $\sim 1 \times 10^{-9}$ erg cm$^{-2}$
s$^{-1}$ (0.8-10 keV) is
consistent with the XIS result of the center observation.
This is also a consistency check that the XIS arf is correctly
estimated.

\begin{table}[!h]
\caption{Fitting results of XIS spectra of cluster offset observations
 with the model of WABS $\times$ APEC.}
\label{offxisfit}
\begin{center}
\begin{tabular}{lllllll} \hline \hline
region    & $N_{\rm H}$  & kT  &  abundance & reduced $\chi^{2}$ &flux(0.8-10 keV)\\ \hline
(Area)  & ($\times 10 ^{21}$ cm$^{-2}$)   & (keV) & (solar) &
 ($\chi^{2}$/d.o.f) & (erg cm$^{-2}$ s$^{-1}$) \\ \hline
offsetA:  & 1.0 fixed  & 5.1 $\pm$ 0.2   & 0.28 $\pm$ 0.07 & 184/195   & 0.85$ \pm 0.02 \times 10^{-9}$ \\ 
offsetB:  & 1.0 fixed  & 5.3 $\pm$ 0.1   & 0.17 $\pm$ 0.03 & 435/440   & 1.47$ \pm 0.01\times 10^{-9}$ \\   
offsetC:  & 1.0 fixed  & 6.3 $ \pm $ 0.2   & 0.21 $\pm$ 0.04 & 327/323 & 1.28$ \pm 0.02\times 10^{-9}$ \\
offsetD:  & 1.0 fixed  & 5.8 $\pm$ 0.5   & 0.32$ \pm $ 0.15 & 46/50    &
 1.03$ \pm 0.04 \times 10^{-9}$ \\ \hline
\end{tabular}
\end{center}
\end{table}

\begin{figure}[!h]
\begin{center}
\rotatebox{-90}{\includegraphics[scale=0.32]{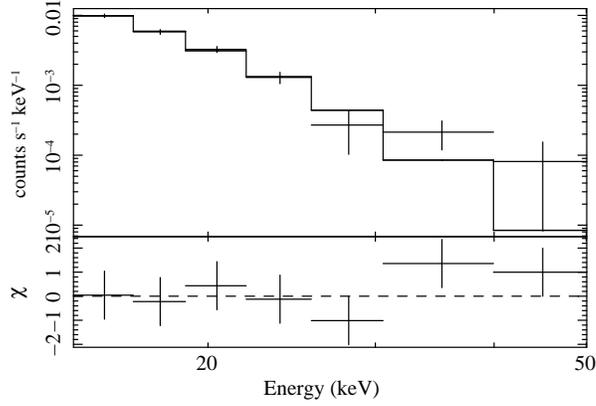}}
\end{center}
\caption{Background-subtracted HXD-PIN spectrum of the
 offset observations with the best-fit single-temperature
 APEC model.}
\label{offpinfig}
\end{figure}

For the PIN, the data are summed over four offset observations, because
the emission from the offset regions is very faint for the PIN and sufficient
statistics cannot be obtained from the individual observation.
We performed spectral fitting by a single APEC model with metal
abundance fixed to 0.25 solar. Best-fit result is shown in Fig
\ref{offpinfig} and Table \ref{offpinpara}. 
The temperature was obtained to be $8 \pm 2$ keV, somewhat
higher than the XIS results.

Next, we performed a joint-fit of the XIS spectrum and the PIN spectrum. 
Here, we summed over the four offset spectra of XIS. We fitted 
spectra with a two-temperature APEC model. We fixed a lower
temperature to 5.5 keV, which is the average temperature in the offset
regions shown
in Table \ref{offxisfit}. A higher temperature is set to be free.
Then, a temperature of 8.7 $^{+3.3}_{-1.8}$ keV is obtained for the
higher temperature. 
Therefore, the XIS and PIN spectra do not need an extremely hot
component in the offset region.

\begin{table}[!h]
\caption{Fitting results of the HXD-PIN spectrum of the cluster offset observations
 with the APEC model}
\label{offpinpara}
\begin{center}
\begin{tabular}{llllll} \hline \hline
kT &abundance & reduced $\chi^{2}$& total flux(15-50 keV)     \\ \hline
(keV) & (solar) & ($\chi^{2}$/d.o.f) & (erg cm$^{-2}$ s$^{-1}$) \\ \hline
 $7.82^{+2.05}_{-1.46}$ & 0.25 fixed & 4.1/5 &
 $1.25^{+0.08}_{-0.36}\times 10^{-10}$ \\ \hline
\end{tabular}
\end{center}
\end{table}

\subsection{Consideration of the PIN collimator response}

\begin{figure}
\begin{center}
\includegraphics*[scale=0.7]{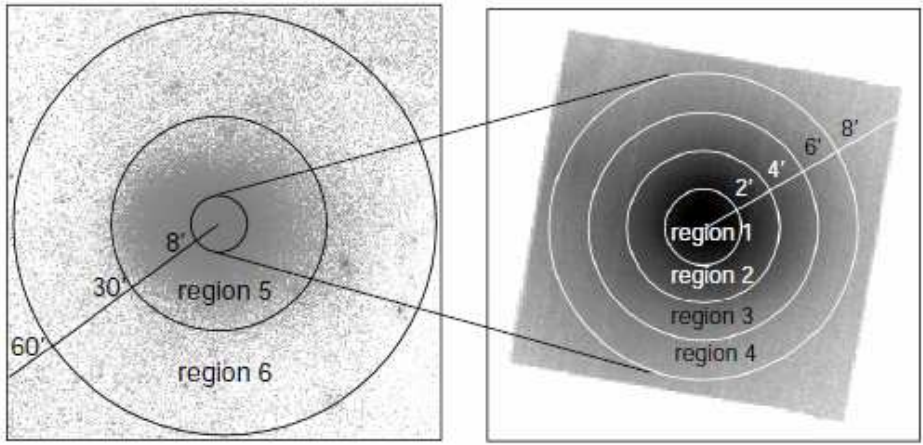}
\end{center}
\caption{Definition of six cluster regions. The radii of circles are 2', 4', 6', 8', 30', 60' from the cluster center.}
\label{regiondefine} 
\end{figure}

In the previous section, 
we roughly obtained the averaged temperatures in the FOV, and they are 
consistent with past observations. 
In this subsection, on the basis of above results, we investigate possible
temperature structure by considering the collimator
response of the PIN correctly.
For that purpose, we implement the Monte-Carlo simulator that reproduces
the PIN spectra of the Perseus cluster by considering the assumed temperature
structure and the collimator angular response of the PIN.
By comparing simulated spectra with actual observed spectra, 
we can find the temperature structure that reproduces 
the observed data the best.
Here, we divide the cluster into six annular regions, 1 - 6, as
shown in Fig \ref{regiondefine}. The definition of region 1--4 is the
same as the previous XIS analysis of the center observation. 
As a first step, assuming the emission model parameters of region 1--4
to be those measured by XIS, we obtain the temperature
of region 5 (8' $<$ r $<$ 30') by using the PIN data of the center
observation. Next, assuming the temperature of region 5 to be the value 
obtained in the first step, as well as region 1-4 temperature, 
we obtain the temperature of
region 6 (30' $<$ r $<$ 60') by using PIN data of the offset observations.

\begin{figure}[!h]
\begin{center}
\begin{minipage}{5cm}
\vspace{0.2cm}
\includegraphics[width=4.3cm,clip]{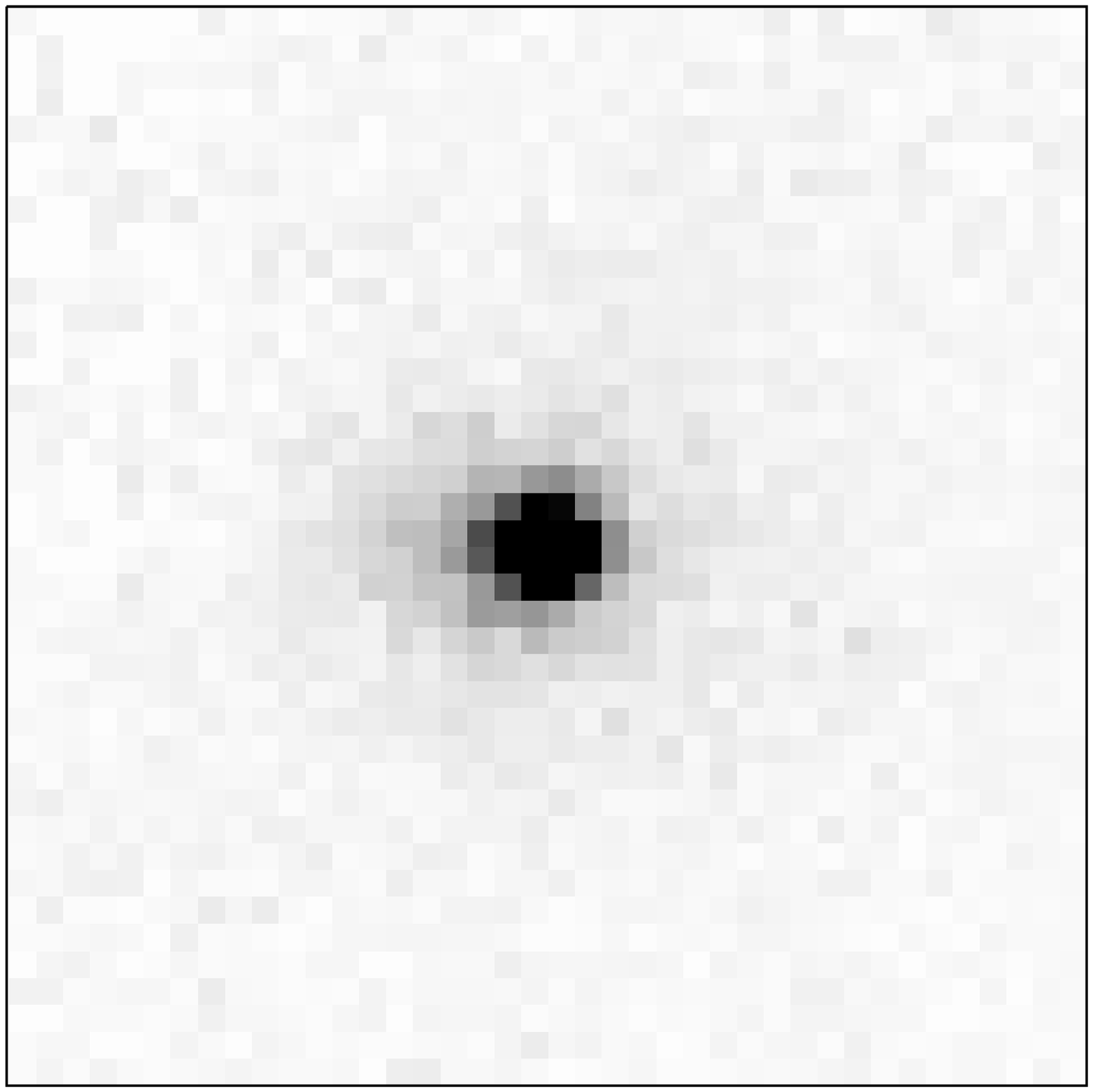}\\
(a)
\end{minipage}
\hspace{0cm}
\begin{minipage}{5cm}
\vspace{0.2cm}
\includegraphics[width=4.3cm,clip]{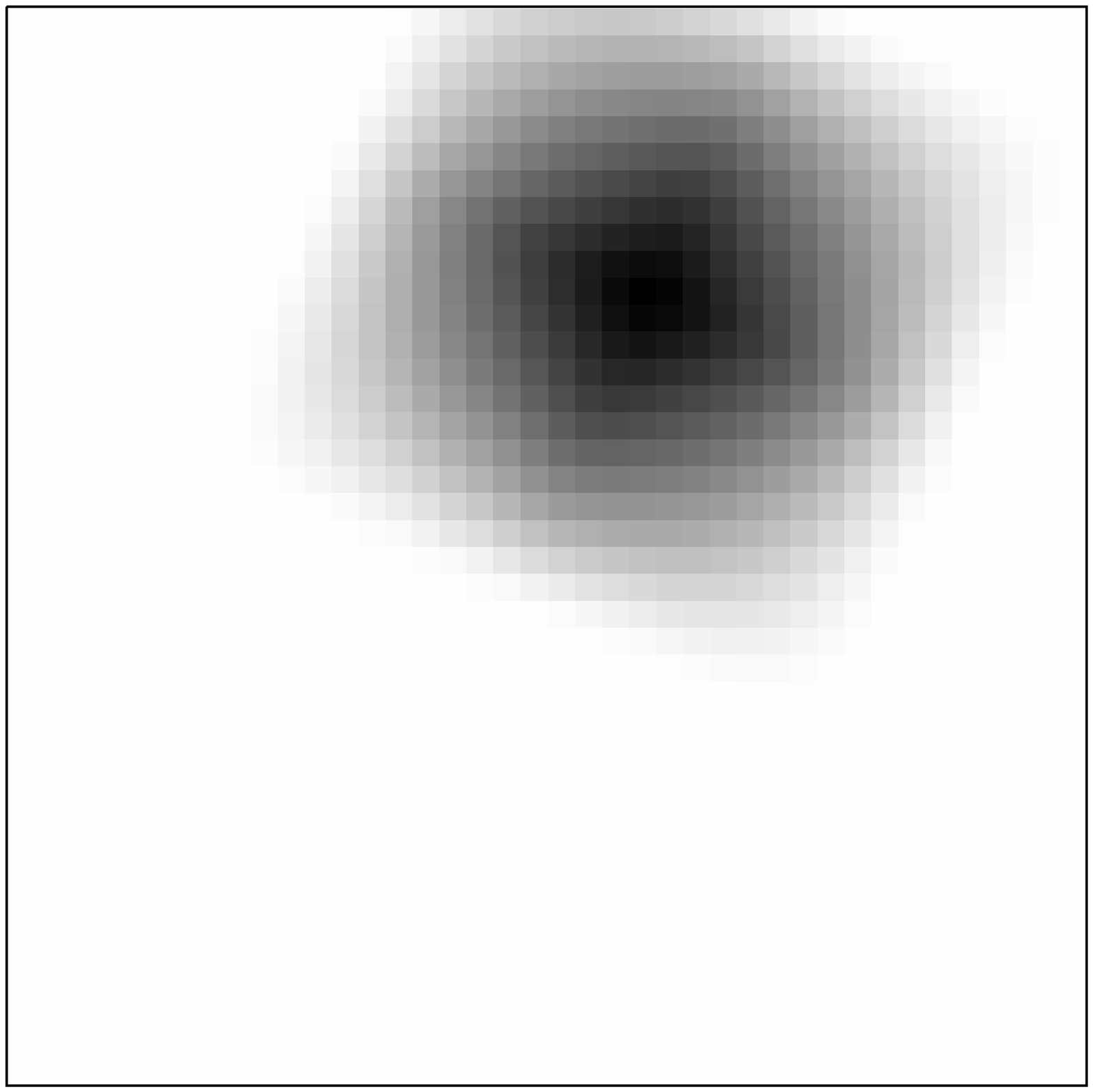}\\
(b)
\end{minipage}
\hspace{0cm}
\begin{minipage}{5cm}
\vspace{0.2cm}
\includegraphics[width=4.3cm,clip]{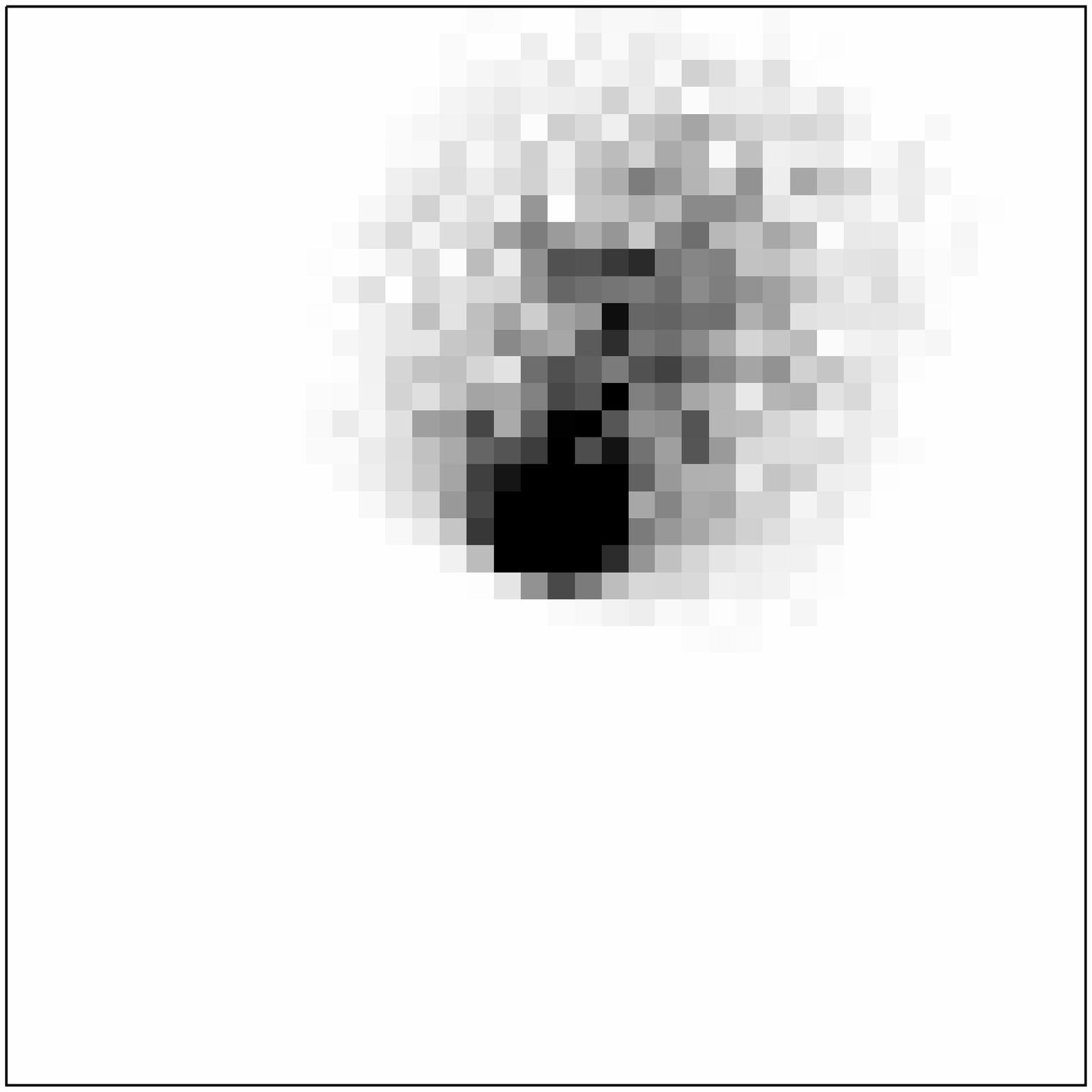}\\
(c)
\end{minipage}
\end{center}
\caption{(a)ROSAT-PSPC image of the Perseus cluster (2 deg $\times$ 2 deg).
 (b)Angular response map for the offset A observation for 30 keV
 photon. (c)Probability map of incident direction of photons in the
 simulation, i.e. map (a) $\times$ map (b).}
\label{labfig9}
\end{figure}

At first, we prepare the emission spectral model with an assumed
temperature, for each region of region 1--6. The spectral model is created
between 0.1 - 100 keV with 256 bins on a linear scale
by the Xspec APEC model. 
In the simulator, we determine a random position of generated photons. 
We used a 2 deg $\times$ 2 deg ROSAT-PSPC image of the
Perseus cluster (Fig \ref{labfig9}a) as a seed photon map. The image is
divided into $40 \times 40$ regions whose size is 3' $\times$ 3'. Then, we
randomly determine an energy of generated photons, following the emission
spectral model we assumed as above. 
Next, thus-generated photons are filtered by the angular response map,
which is created by the FTOOL hxdarfgen for each observation.
An example of the angular response
map for 30 keV photons is shown in Fig \ref{labfig9}b. The transmission
efficiency of the
collimator is almost constant below an energy of 50 keV. 
The number of simulated photons is $10^{7}$ for each observation.
Output is a spectrum to which only
the angular response of the PIN detector is applied.
The simulated spectrum is read into Xspec as a model, and compared
with the observed spectrum by spectral fitting, 
after the rmf is applied to the model. 
Here, only the normalization of the model is a free parameter.

\begin{figure}[!h]
\begin{center}
\begin{minipage}{8.0cm}
\includegraphics[scale=0.65]{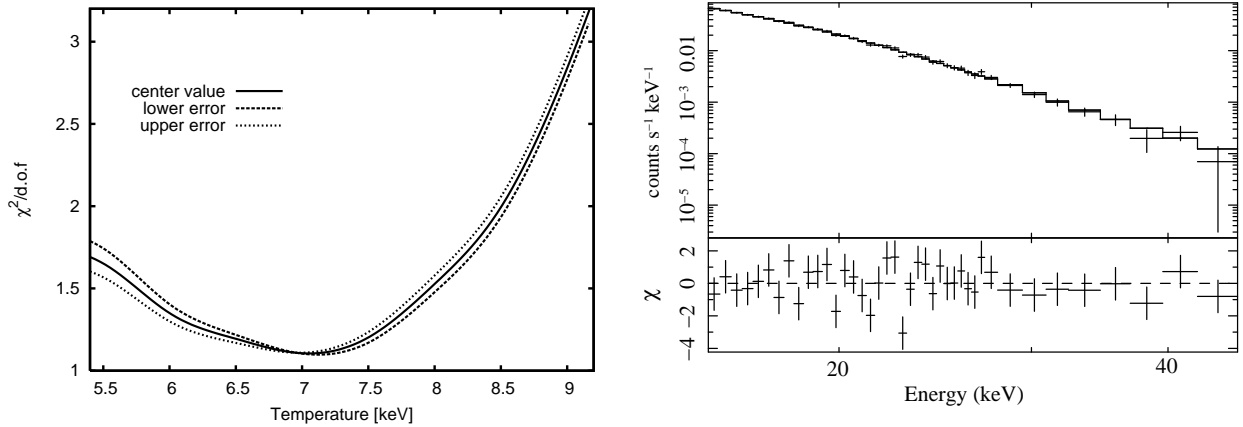}
\end{minipage}
\hspace{0.2cm}
\begin{minipage}{8.0cm}
\rotatebox{-90}{\includegraphics[scale=0.33]{./paper_fig10b.eps}}\\
\end{minipage}
\end{center}
\caption{(left) Reduced $\chi^{2}$ of the difference between the actual
 spectrum and simulated spectrum for region 5. (right)
 Background-subtracted HXD-PIN spectrum of the center observation with
 the best-fit model (kT of region 5 is 7.0 keV).}
\label{centerchi}
\end{figure}

First, we obtain the temperature of region 5, 
using the PIN data of the center
observation. The emission models of region 1-4 are based on the XIS
results in Tables \ref{hpara} and \ref{cx}. 
The temperature of region 5 is tested 
from 5 keV to 12 keV with a step of 0.5 keV, with metal abundance fixed 
to 0.35 solar. The emission model of region 6 is still
unknown, and thus assumed to be the same as that of region 5. 
The fraction of incident photons from region 6 is approximately 1\% of 
all photons, and thus their contribution is negligible.
Moreover, the contamination of point sources in the FOV, for
example IC 310 shown in the XIS image of offset B observation, is less
than 1\% of all photons, and thus their contribution is also negligible.
We compare each simulated spectrum with the
observed spectrum by adjusting the model
normalization. 
The $\chi^{2}$ distribution for the temperature of region 5 is shown in Fig \ref{centerchi}
(left). Here, the uncertainties
of the XIS model parameters in region 1--4 are taken into account and error
ranges are shown in the figure as dashed lines.  
The best-fit temperature of region 5 is 7 keV, and the error range is 
6.6--7.4 keV at 90\% confidence level. 
Fig \ref{centerchi} (right) shows the best-fit model and spectrum.
The fraction of photon flux from region 1--6 in the best-fit
simulated spectrum is 6.2\%, 17.0\%, 25.5\%, 15.8\%, 34.4\%, 1.1\%.

Next, to obtain the temperature of region 6 (30' $<$ r $<$ 60'),  
we adopt the temperature of 7 keV for region 5, which is as obtained above,
in addition to the emission model of region 1--4.
The temperature of region 6 is tested from 5 keV to 15 keV with a step
of 1 keV, with metal abundance fixed to 0.25 solar.
We investigated the averaged-temperature of the four offset
regions, by using the PIN spectrum summed over the four observations as
in \S3.1 and comparing it with the simulated spectrum.
Results are shown in Fig \ref{addpower}.
The best-fit temperature of region 6 is $\sim$9.0 keV. The upper limit
is $\sim$ 14 keV at the 90\% confidence level, while the lower limit is not
well constrained.
The fraction of photon flux from region 1--6 in the best-fit simulated
spectrum is 2.9\%, 7.8\%, 14.3\%, 11.0\%, 40.2\%, 23.8\%.
The number of photons from region 5 and 6 is more than 60
\%, and the contribution from the central region is not so large 
in the offset observation.
In fact, the temperature in region 6 does not depend so much on the 
power-law flux of NGC 1275. 
We considered the systematic errors due to the uncertainty of the NXB
(Non X-ray Background) of HXD-PIN. The HXD-team announces that NXB
models released by them have $\pm$ 3\% uncertainty (Fukazawa et al. 2009). 
When we decreased the NXB by 3\%, an upper limit of the temperature becomes
7.9 keV for region 5 and 19 keV for region 6.

\begin{figure}[!h]
\begin{center}
\begin{minipage}{8.0cm}
\vspace{0cm}
\includegraphics[scale=0.65]{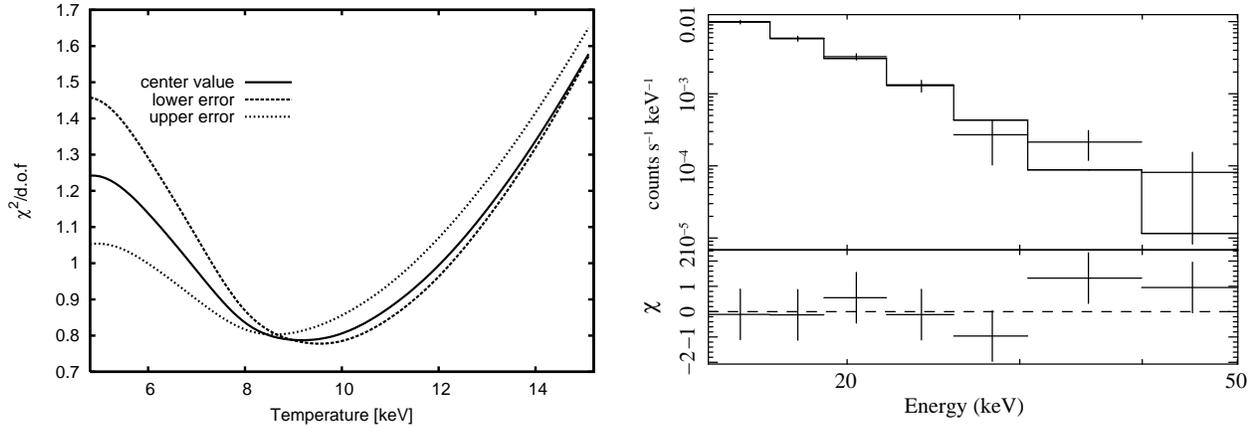}
\end{minipage}
\hspace{0.2cm}
\begin{minipage}{8.0cm}
\rotatebox{-90}{\includegraphics[scale=0.33]{./paper_fig11b.eps}}
\end{minipage}
\end{center}
\caption{(left) Reduced $\chi^{2}$ of the difference between the actual
 spectrum and simulated spectrum for region 6. (right)
 Background-subtracted HXD-PIN spectrum of the offset observation with
 the best-fit model (kT of region 6 is 9.0 keV).}
\label{addpower}
\end{figure}

\subsection{Constraints to non-thermal emission from the cluster}

Here, we constrain the non-thermal emission hidden
by the thermal emission in the offset region of the Perseus cluster. 
First, we performed
a naive spectral fitting for the offset PIN spectrum as \S3.2 (without
considering the PIN collimator response), with a
model consisting of a single
temperature APEC plus POWERLAW. Here, the metal abundance was fixed to 0.25
solar and the photon index of the POWERLAW model was fixed to 2.0. The
best-fit spectrum
is shown in Fig \ref{addpower} (left). A POWERLAW component with a
flux of 5.6 $\times 10^{-11}$ erg cm$^{-2}$ s$^{-1}$ (15 - 50 keV) was
marginally required, but the significance is low.

Next, we performed the spectral fitting by considering the PIN
collimator response as described in \S3.3. We applied a simulated model where
the temperature of region 6 is assumed to be 7 keV, 
the same as for region 5. 
In this case, a more realistic upper limit is obtained than the
naive analysis because the temperature gradient at the center region and
the hard emission from NGC 1275 are taken into account in this
analysis. 
Actually, a tighter upper limit of 4.4 $\times 10^{-12}$ erg cm$^{-2}$
s$^{-1}$ is obtained as shown in Fig \ref{addpower} (right) and Table
\ref{addpowertable} (bottom).

\begin{figure}[!h]
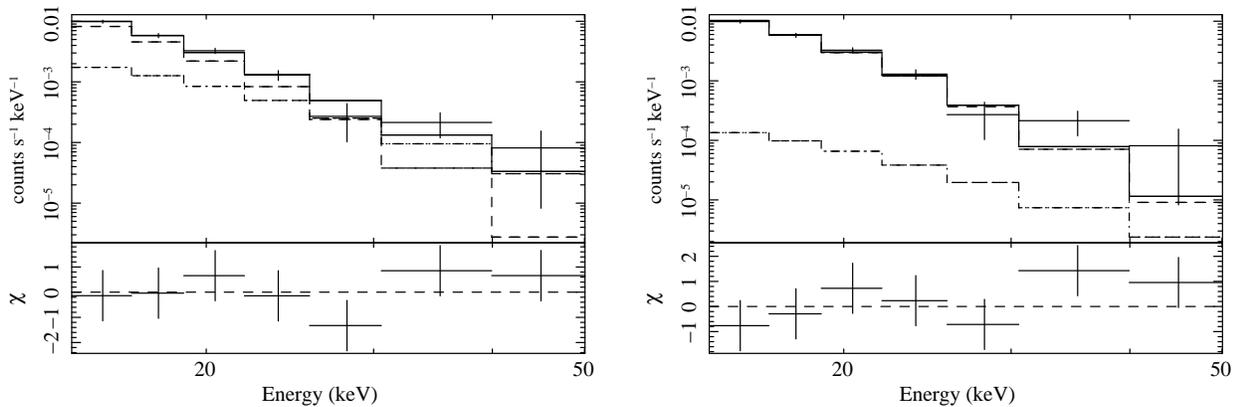

\begin{center}
\begin{minipage}{8.0cm}
\rotatebox{-90}{\includegraphics[scale=0.32]{./paper_fig12a.eps}}
\end{minipage}
\hspace{0.2cm}
\begin{minipage}{8.0cm}
\rotatebox{-90}{\includegraphics[scale=0.32]{./paper_fig12b.eps}}
\end{minipage}
\end{center}
\caption{Background-subtracted HXD-PIN spectra of the offset
 observations with the best-fit model of the single temperature APEC plus
 POWERLAW model (left), with the best-fit model of the simulated thermal
 model plus POWERLAW model (right). Dashed lines in the right upper
 panels show the APEC (left)/simulated thermal (right) and POWERLAW contributions.}
\label{addpower}
\end{figure}

\begin{table}[!h]
\caption{Fitting results of the PIN spectra of the offset observations.}
\label{addpowertable}
\begin{center}
\hspace*{-1.0 cm}
\footnotesize
\begin{tabular}{lllllll} \hline \hline
photon index & reduced $\chi^{2}$& PL flux(15-50 keV) & total flux(15-50
 keV)\\ \hline
(-) & ($\chi^{2}$/d.o.f)& (erg cm$^{-2}$ s$^{-1}$) & (erg cm$^{-2}$ s$^{-1}$) \\ \hline
2.0 fixed & 4.7/5 & $4.38 \times 10^{-12}$(upper limit) & $1.22 \times 10 ^{-10}$ \\ \hline
\end{tabular}
\end{center}
\end{table}

\section{Discussion}

\subsection{Temperature structure}

\begin{figure}[!ht]
\begin{center}
\rotatebox{-90}{\includegraphics*[scale=0.35]{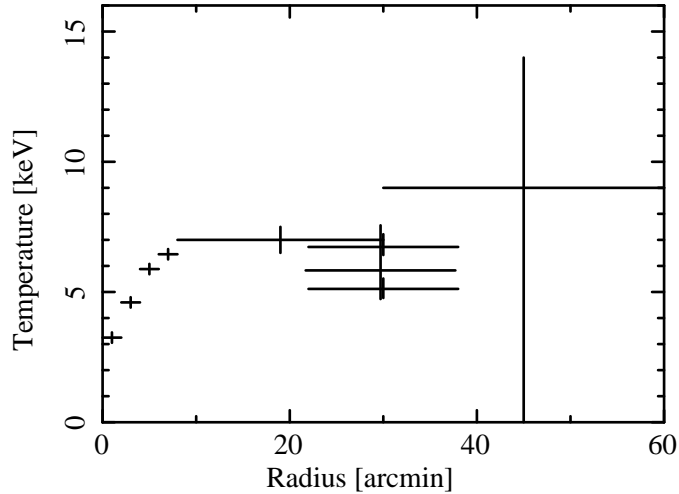}}
\end{center}
\caption{Radial temperature profile of the Perseus cluster, obtained in
 this work. The first four crosses from the center are
 obtained with the XIS in \S3.1 and \S3.2, crosses around 20' and
 45' are obtained with the PIN in
 \S3.3. Furthermore, temperatures obtained from XIS offset observations
 are shown at 30' for the offset C, D, A in order of
 temperature decreasing. The temperature of offset B containing emission
 from IC 310 is not shown here. Error bars in the figure present
90\% confidence level.}
\label{simtemp}
\end{figure}

The radial temperature profile obtained in this work is shown
in Fig \ref{simtemp}. Error bars in the figure present the 
90\% confidence level, though systematic errors due to the NXB uncertainty of
the HXD-PIN are not shown here.
Distinct features of the temperature structure are a steep decline toward
the center and a relatively flat profile in the outer region.
XIS results in the offset regions give a lower temperature 
than the PIN analysis. 
This is possibly because the XIS is sensitive to a lower 
temperature component, if two different temperature (hot and cool) 
components exists at different regions along the same line of sight. 
Therefore, the results of
the XIS and the PIN do not necessarily conflict with each other.
Most importantly, we found that the upper limit of the temperature of
the offset region, where hot components seem to exist, is 
at most 14 keV ($<$ 19 keV within the systematic errors). Considering
that the XIS
spectra do not require the hot
component, it seems that very hot gas as reported in RXJ
1347.5-1145 (Ota et al. 2008) or A 3667 (Nakazawa et al. 2009) does not
exist in the Perseus cluster.
In this analysis, we could not study temperature fluctuations 
due to the low data statistics of the offset observations. But
the average temperature in each offset region is not inconsistent with the
results of ASCA (Furusho et al. 2001).

Temperature structure with a negative gradient towards the center and a
roughly flat external plateau is typical for cooling flow
type clusters (Pointecouteau et al. 2005). These clusters,
showing a strong cooling core, are thought to be gravitationally
relaxed. Actually, most colliding-type clusters do not show such a cooling
core. Therefore, we infer that the main body of the
Perseus cluster has been already relaxed, and has not recently
experienced a violent cluster merger, which greatly influences the
temperature structure of the cluster. 
On the other hand, ROSAT and other X-ray observatories found an excess
of the surface brightness at 20' east from the cluster center (Ettori et al.
1998). Moreover, ASCA found a temperature
drop at the corresponding region (Furusho et al. 2001). Therefore, it is
implied that a small-scale substructure, such as a small galaxy cluster or
group with cool gas, is currently colliding with the main body of the
Perseus cluster.

We consider two reasons why extremely hot gas does not exist in the 
Perseus cluster, even
though some evidences of the cluster merger were found as described above.
The first reason is that the present merging is in the latest
phase, that is, extremely hot gas heated in the long past has already cooled
down. The second reason is that heating condition is not satisfied in
the current merging. In the case of RXJ 1347.5-1145 (Ota et al. 2008),
though it depends on various cluster parameters, extremely hot gas
seems to take approximately 0.5 $\times$ 10$^{9}$ yrs to cool down by
radiation. On the other hand, the cold substructure in the
Perseus cluster must have been created in recent 10$^{9}$ yr, 
considering the diffusion time of the ICM (Furusho et al. 2001), and
this is roughly comparable to the cooling time of extremely hot gas.
Accordingly, the picture that only the extremely hot gas cooled down without
diffusion of the cold gas is not natural. Therefore,  
the first reason, that a violent cluster merger occurred in the
long past and extremely hot gas has already cooled down, is unlikely.
Hence, the second reason is preferred.
According to Kitayama et al. (2004) and Ota et
al. (2008) with help of Takizawa (1999), 
a high velocity ($\Delta \nu \sim$ 4500 km
s$^{-1}$) collision of two massive (5 $\times$ 10$^{14}$ M$_{\odot}$)
clusters is needed to heat the ICM up to 20 keV or
higher for RXJ 1347.5-1145. 
In the case of the Perseus cluster, the total mass, including the cold
substructure, is 1.2 $\times$ 10$^{15}$M$_{\odot}$, according to Ettori et
al. (1998). However, the mass of the subcluster is much less than the total
mass, probably by one or two orders of magnitude. 
Then, it may be difficult to heat
up the ICM to a high temperature, if the mass difference of two merging
clusters is very large. Actually, according to Takizawa et al. (1999),
two-cluster merging, with a mass ratio of 4, 
could heat up the ICM to at most $\sim$ 13 keV.
Therefore, it is inferred that the current Perseus cluster is already
grown up enough, and not in a violent merger phase.
Though it is just speculation, a history of the cluster formation may be
divided into two phases. 
In the early phase, clusters or groups with almost the
same mass repeat violent mergers and grow up. In the latest phase, a grown-up
large cluster cannibalizes small-scale galaxies clusters or groups. 
Therefore, comparison between the spatial distribution of merging
clusters and relaxed clusters is interesting to understand the evolution
of the galaxy clusters and the large-scale structure of the universe.

\subsection{Non-thermal emission from the Perseus cluster}

In our analysis, we obtained the upper limit flux of the non-thermal
emission as 4.4 $\times$ 10$^{-12}$ erg cm$^{-2}$ s
$^{-1}$ (15 - 50 keV) at the cluster offset region. 
So far, the Perseus cluster has been often observed by X-ray
satellites, such as ASCA, XMM-Newton,
Chandra, Swift/BAT (Ajello et al. 2009) and so on. 
However the non-thermal emission from the cluster itself has not been 
discussed well, because the thermal emission from the
cluster core is very bright 
and furthermore the AGN emission from NGC 1275 must be considered.
In this paper, we performed observations of cluster offset regions, by
reducing the bright emission from the cluster center.
In addition, we considered the contribution of the AGN emission from
NGC 1275. 
Therefore, this may be the first robust result of an upper limit of
non-thermal emission from the Perseus cluster. 
Assuming the distance to the Perseus cluster to be 75 Mpc, the corresponding
upper limit luminosity is 3.0 $\times$ 10$^{42}$ erg s$^{-1}$. 
This limit is tighter than those of other clusters
observed with Suzaku HXD-PIN, such as A 3376 (Kawano et al. 2009) and
A3667 (Nakazawa et al. 2009), owing to the proximity of the Perseus
cluster. Considering no report of radio detection of the largely
extended synchrotron emission like Coma cluster, there is yet no
evidence of particle acceleration in the Perseus
cluster. This result also supports that the Perseus
cluster is not in a violent cluster merger phase, as inferred from
the ICM temperature studies. \\

The authors thank Dr. Jelle Kaastra for careful reading and many useful
comments.
The authors also thank the Suzaku team for development of hardware/software
and operation. SN is supported by Research Fellowships of the Japan
Society for the Promotion of Science for Young Scientists.


\end{document}